\documentclass[12pt,preprint]{aastex}

\shorttitle{The Flare/CME Model of Gamma-ray Bursts}
\shortauthors{aoki et al.}

\begin{document}


\title{The Log-normal Distributions of Physical Quantities of Flare associated Coronal Mass Ejections (CMEs), and Flare/CME Model of Gamma-ray Bursts}


\author{Seiichiro I. Aoki \altaffilmark{1,2,3}}
\email{aoki@kwasan.kyoto-u.ac.jp}

\author{Seiji Yashiro \altaffilmark{4}}
\email{yashiro@cdaw.gsfc.nasa.gov} 
\and
\author{Kazunari Shibata \altaffilmark{1}}
\email{shibata@kwasan.kyoto-u.ac.jp}
%% are identified by the \altaffilmark after each name. Specify alternate

\altaffiltext{1}{Kwasan and Hida Observatories, Kyoto University, Yamashina, Kyoto 607-8471, Japan}
\altaffiltext{2}{Department of Astronomy, Faculty of Science, University of Tokyo, 7-3-1 Hongo, Bunkyo-ku, Tokyo 113-0033, Japan}
\altaffiltext{3}{Solar Division, National Astronomical Observatory, 2-4-1 Osawa, Mitaka, Tokyo, 181-8588, Japan}
\altaffiltext{4}{Center for Solar Physics and Space Weather, Catholic University of America, Washington, DC 20064}


\begin{abstract}
We investigated the statistical distributions of physical quantities of
solar flares and associated coronal mass ejections (CMEs). We found
that the distributions of the X-ray peak fluxes of CME-related flares,
their time intervals,  and speeds of associated CMEs  are in good agreement
with log-normal distributions.  One possible interpretation of this is
that only large energetic mass ejections can escape from the solar corona,
which become CMEs in the interplanetary space.
This ``filtering effect'' may be the reason why
statistical distributions of some physical quantities 
are similar to log-normal distribution.
It is known that the distributions of
several physical quantities of gamma-ray bursts (GRBs) are also
log-normal distributions, and that the time variations of
gamma-ray intensity in GRBs are very similar to those of hard X-rays of
solar flares.   On the basis of these similarities and
physical consideration of magnetohydrodynamic properties
of an accretion disk around a black hole,
which is supposed to be situated in the central engine of a GRB,
we propose a new model of the central engine of GRBs,
the {\it flare/CME model},
in which GRBs are formed by nonsteady jets consisting of
intermittent ejections of mass (like solar CMEs) associated with
reconnection events (like solar flares) in the accretion disk corona. 
Such nonsteady MHD jets would produce many shocks 
as well as local reconnection events 
far from the central engine. In this model, only
large energetic mass ejections can escape from the 
accretion disk corona, so that statistical distributions of some physical
quantities are similar to log-normal distributions.
\end{abstract}




\section{INTRODUCTION}
Gamma-ray bursts (GRBs) are the most luminous objects in the universe, 
and known to be highly  time-variable. 
Solar flares also show highly time variable feature, and are
similar to GRBs from the viewpoint of X-ray and gamma-ray light curves.
In fact, the power spectra of the light curves of GRBs 
are power-law distribution (Beloborodov, Stern, \& Svensson 2000) 
and similar to those 
in solar X-ray emissions (Ueno et al. 1997). 
Hence, it has often been argued that
the physical process of GRBs may be similar to that of solar flares. 
However, there is one important difference in the statistical distributions 
of the physical quantities of GRBs and solar flares. That is, some
quantities such as total durations,  interpulse and interspike time 
intervals (McBreen et al. 1994), peak fluences and peak intervals 
(Li \& Fenimore 1996), 
pulse durations (Nakar \& Piran 2002), and break energies (Preece et al. 2000)
show the log-normal distributions in the case of GRBs, 
while those of solar flares 
such as peak count rates (energy), waiting times, and total durations are 
power-law distributions (Dennis 1985; Hudson 1991; Pearce, Rowe, \& Yeung 
1993; Crosby, Aschwanden, \& Dennis 1993).
What is the physical reason of this difference? 
Does this imply that the underlying physics of GRBs is fundamentally 
different from that of solar flares ?  

The purpose of this paper is to explore the physical meaning of
this difference using the actual data of solar flares and associated
coronal mass ejections (CMEs).  On the basis of this  comparative
study between solar flares/CMEs and GRBs, we shall further try
to understand the nature of the central engine of GRBs. 
Here,  we note  that the difference in the statistical 
distributions of physical quantities between solar flares and GRBs 
may be due to the difference in photon-emitting regions. 
GRBs are believed to occur on 
the relativistic outflow ejected from the central engine of a GRB 
(e.g., Goodman 1986; Paczy\'{n}ski 1986). Moreover, the outflow from the 
central engine can be collimated into a jet 
(Frail et al. 2001; Panaitescu \& Kumar 2001); 
it is likely that the GRBs are 
due to the relativistic jets ejected from the central engine.
On the other hand, the emissions associated with solar flares are produced 
in the low corona of the Sun. Hence, when we compare the distributions of 
the physical quantities of GRBs with those of solar flares, we must not 
take solar flares but the phenomena associated with mass outflow from the Sun,
 which are CMEs. 

Recent solar observations have revealed that magnetic reconnection is a key
mechanism for the energy release in solar flares (e.g., Tsuneta et al.
1992, Masuda et al. 1994), and that 
mass ejections (often called plasmoid or flux rope ejections) are 
ubiquitous in solar flares, not only in large scale flares, 
but also in small scale flares
(e.g., see Shibata 1999 for a review). 
On the basis of these new observations,
Shibata et al. (1995) have proposed the plasmoid-induced-reconnection 
model, which is a unified model for solar flares,
from largest CME-related flares to small flares, even 
microflares and nanoflares. The important point in this model is that
the reconnection is closely coupled to mass (plasmoid) ejections and 
hence the mass ejection occurs in all reconnection processes (flares)
(Shibata 1999).
The difference in apparent morphology in various flares 
is largely a result of difference in relative size and/or energy of flares 
compared with those of ambient magnetic structure:
that is, if the size of the flare is small and less energetic,
it is surrounded by strong magnetic field and so mass ejection is soon 
stopped or trapped, and cannot be escaped from the corona. In this case,
flares look like a simple loop in soft X-rays, and 
no CME is associated with the flare. 
However, if the size of the flare is large and energetic, the ejection
is escaped from the corona and can become CMEs.
In this case, flares look like a cusp-shaped loop in soft X-rays. 
Hence this scenario predicts that there is a typical 
value for some quantities (total energy, size, duration, or any
quantity reflecting the condition for escape from the corona, i.e., 
the condition for CME), which
lead to statistical distribution similar to  log-normal distribution. 
Thus, in the present paper, in order to test this prediction,
we investigated the distributions of the physical quantities of 
the solar flares associated with CMEs, which we simply call 
"CME-related solar flares".

A log-normal distribution is the Gaussian (or normal) distribution 
on the plot whose horizontal axis is taken in logarithmic scale and 
vertical axis is presented in linear scale.
A log-normal distribution is defined as follows 
(Aitchison \& Brown 1957; Crow \& Shimizu 1988):
\begin{eqnarray}
f(x) = \left \{  \begin{array}{lr}
\frac{1}{\sqrt{2 \pi} \sigma} \exp{\left [ -\frac{(\log{x}-\mu)^2}{2 \sigma^2} \right] }  & (x>0),
\\
0 & (x \leq 0),
\end{array} \right.
\end{eqnarray}
where $f(x)$ is the probability density function for $x$, and $\mu$ and 
$\sigma^2$ are the sample mean and variance of $\log x$, respectively. 
In general, 
the origin of a log-normal distribution has been explained by the statistical effect of the production of the physical quantities, that is, the effect of the central limit theorem. Ioka \& Nakamura (2002) have shown that the origin of the log-normal distribution of the physical quantities of GRBs can be explained in the same manner. However, the log-normal distribution of total durations has not been explained yet by them, so that their explanation is not complete one.
Therefore, we take attention to the characteristic of a log-normal distribution that there is a typical value in the distribution.
If there is no typical value, the distribution becomes a scale-free one like a power-law distribution.
We think the existence of a typical value in the distribution of the physical quantities is possibly due to the physical process.

In this Letter, we describe the method of the data analysis 
in section 2,  
and presents the results in section 3. On the basis of these data
analysis of the CME-related solar flares, we shall discuss comparative
study of solar flares/CMEs and GRBs, and propose a "flare/CME model" 
for the central engine of GRBs in section 4. 

\section{DATA ANALYSIS}
We investigated the distributions of the X-ray peak fluxes, time intervals, and CME speeds of the CME-related solar flares. The X-ray peak flux and time interval of the CME-related solar flares correspond to the peak fluence and peak interval (interpeak and interspike time interval) of GRBs, respectively. 
The data of the CME-related solar flares are generated  
by the comparison between solar flares and CMEs as follows.
At first, we prepare CME data from the SOHO/LASCO CME CATALOG 
(http://cdaw.gsfc.nasa.gov/CME\_list/)
for the period from 1996 October 19 to 2001 December 26,
and solar flare data from the Solar Geophysical Data (SGD) for the same period.
Second, we calculate the leaving time of CMEs from the surface of the Sun ($t_l$) by interpolating from the detected time of the CMEs, assuming that the speed of the CME remains constant from the solar surface to the detected position. 
The interpolation is necessary because CMEs cannot be detected at the solar surface but at several solar radii from the Sun by the detection limit of SOHO/LASCO. 
Then we compare the starting time ($t_s$) and peak time ($t_p$) of a solar flare with 
the leaving time of a CME. If the leaving time of the CME is between the starting time and peak time of a solar flare ($t_s  \le t_l \le t_p$), we identify the solar flare as a CME-related solar flare.
Along with this rule, we select 254 samples of the CME-related solar flares.
However, when we investigate the time interval distribution of the CME-related solar flares, 
we pick up 165 samples whose angular widths are larger than $60$ degree from the data of the CME-related solar flares. 
We take this selection rule, because not all narrow CMEs can be identified by the detection limit.

\section{RESULTS}
We found that the distribution of the X-ray peak fluxes, time intervals, and CME speeds
of the CME-related solar flares are in good agreement with log-normal distributions (Fig. 1). In Figure 1, the dashed curves show the best-fitting log-normal distributions.
The X-ray peak flux at the peak position of the number of the CME-related solar flares is $\sim 6.81 \times 10^{-6} \mathrm{W/cm^2}$ (Fig. 1a). The log-normal distribution has a typical value, so that CMEs are associated with energetic flares. The distribution with X-ray peak fluxes weaker than the peak position is small, so that the distribution of X-ray peak fluxes is deviated from the log-normal distribution. This is due to the threshold of detecting solar flares, which depends on solar activity. If the activity of the Sun is high, the threshold is large and the distribution in weak X-ray peak flux becomes small. However, the maximum threshold is $\sim 10^{-6} \mathrm{W/cm^2}$, so that the existence of the peak position is not due to the threshold. The time interval at the peak position of the number of the CME-related solar flares is $\sim 118$ hours (Fig. 1b). The CME speed at the peak position of the number of the CME-related solar flares is $\sim 605$ km/s (Fig. 1c). 

We note here that not all narrow CMEs are listed in the SOHO/LASCO CME CATALOG, since there is possibility that some faint and narrow CMEs were missed; it is difficult to identify narrow CMEs which occurred just after large CMEs. As for solar flares, weak ones are not included by the detection limit of GOES, neither. 
In addition, when a solar flare occurs on the hidden side of the Sun, 
it cannot be detected. Thus, our data are not complete ones. 
Nevertheless, we believe the distributions would not be different 
from log-normal distributions even when the quality of the data 
will be improved in the future. 

\section{DISCUSSION}
The distributions of the physical quantities of the CME-related solar flares and those of GRBs are log-normal distributions, whereas those of solar flares are power-law distributions. Cygnus X-1, one of the black hole candidates, also shows a highly time-variable feature (Miyamoto el al. 1992) like GRBs and solar flares.
It has been found that the distributions of the X-ray peak intensity and peak interval of Cygnus X-1 
show nearly exponential distributions (Negoro 1995; Negoro et al. 1995). An exponential distribution does not have a typical value like a power-law distribution.
The X-ray emissions are considered to come from the accretion disk in 
Cygnus X-1. However, when the shots with large peaks are 
picked up, which may be related to outflows (or jets) in Cygnus X-1, the distributions of the X-ray peak intensity and peak interval become log-normal distributions (Negoro \& Mineshige 2002). 

Therefore, the distributions of the physical quantities associated with 
mass outflows show log-normal distributions, whereas those related to 
surface emissions show scale-free distributions such as power-law and 
exponential distributions. Hence, the log-normal distribution of the physical quantities, which has a typical value, is possibly resulted from the filtering effect on mass ejections. On the basis of the similarity of characteristics among GRBs, solar flares, and Cygnus X-1, we suggest that GRBs are due to the mass outflow associated with flare-like process. In fact, a system of a stellar mass black hole and an accretion disk around it can be formed in the central engine of a GRB in the standard models such as the collapsar model (Woosley 1993; Paczy\'{n}ski 1998; Fryer, Woosley, \& Hartmann 1999), and the NS-NS (BH) merger model (Paczy\'{n}ski 1986; Goodman 1986; Eichler et al. 1989; M\'{e}sz\'{a}ros \& Rees 1997). The accretion disk has magnetic field with $\beta \sim 10 - 100$, where $\beta$ is the ratio
of gas pressure to magnetic pressure (e.g., Machida and Matsumoto 2003). 
This magnetic field is strong enough
to generate lots of magnetic activities in the accretion disk corona.
There is shear between the inner edge of the disk and the black hole. The shear is also in the disk because of the differential rotation of the disk. Therefore, the magnetic loop formed on the disk is twisted and stretched, and magnetic reconnection occurs. In previous studies, Narayan, Paczy\'{n}ski, \& Piran (1992) have suggested that the magnetic flares can be generated by Parker instabilities on the differentially rotating disk in the central engine of a GRB. Recently Klu\'{z}niak \& Ruderman (1998) have proposed that magnetic field is strengthened  in the differentially rotating collapsed object in the central engine of a GRB and is pushed up through the surface of the object by magnetic buoyancy. Hence, there can be many flares similar to solar flares on the disk in the central engine of a GRB.

Altogether we suggest the flare/CME model of the central engine of a GRB by the analogy to solar flares and Cygnus X-1 (Fig. 2). In our model, GRBs and afterglows are produced as follows. 
There are a lot of flares on an accretion disk in the central engine of a GRB, and many plasmoids are produced and accelerated via magnetic reconnection.
If the scale of a magnetic loop is small, a plasmoid cannot gain large energy enough to escape from the magnetosphere of the central engines, and a flare associated with the reconnection is not energetic. Only the plasmoids associated with energetic flares can be escaped. In fact, Hayashi, Shibata, \& Matsumoto (1996) have performed MHD simulations of protostellar jets, and found that magnetic reconnection occurs near the central object, where magnetic loops are strengthened and twisted by shear between the inner edge of the accretion disk and the surface of the central object, and then plasmoids are produced and ejected via the reconnection.
Various flares with different energy scales occur repeatedly on the disk in the central engine, so that a lot of plasmoids with various speeds are ejected repeatedly from the central engine. If the preceding plasmoid is slower than the following one, they will collide with each other.
The structure of the ejected plasmoids would be similar to that of a spheromak, which is unstable to tilting instability (e.g., Hayashi \& Sato 1984). Hence, there are various angles between the plasmoid (spheromak) axis and  
the ejected direction.  When the tangent magnetic field of an ejected plasmoid is opposite to that of other one, magnetic reconnections occur. 
Therefore, all magnetic energy in the ejected plasmoids contained as Poynting flux can be converted to the kinetic and internal energy of plasmas. In addition, a lot of MHD shocks occur in the reconnection process, 
which is convenient for acceleration of high energy particles. 
These may correspond to internal shocks in the internal shock model  (Rees \& M\'{e}sz\'{a}ros 1994). In fact, the collisions between two CMEs, 
and shock formation and electron accelerations due to the collisions were 
observed by radio observation in actual CMEs (Gopalswamy et al. 2001). 
Hence, the collisions between the escaped plasmoids from the central engine of a GRB can occur and internal shocks are formed.
Moreover, the merged shell is produced via the collision. If it propagates into interstellar matters, the external shocks can be formed, which are observed as afterglows. 

The merit of our model to the fireball model (see Piran 1999 and M\'{e}sz\'{a}ros 2002 for recent reviews) is that we can avoid baryon contamination. There are a lot of baryonic plasmas near the central engine of a GRB, so that it is difficult to accelerate the plasmas to relativistic speed near the central engine.
 On the other hand, in our model magnetic energy (Poynting flux) is dominated
near the central engine. 
The magnetic energy is carried by ejected plasmoids and converted to the 
kinetic and internal energy of plasmas by magnetic reconnection 
at the distant place from the central engine.

Recently, several authors have proposed the model of the central engine 
of a GRB via magnetic reconnection (Spruit, Diagne, \& Drenkhahn 2001; 
Drenkhahn \& Spruit 2002; Lyutikov \& Blandford 2002). In their models, 
the magnetic energy is carried to the distant place from the central 
engine as Poynting flux, and reconnection occurs. 
If the statistical distribution of sub-bursts are determined by the 
reconnection  in the outflow at the distant place from the central engine,
they may not show log-normal distributions, since the reconnection itself
generally tend to produce power-law (or scale free) distribution.
Thus, it would be difficult to explain the log-normal distributions of 
the physical quantities of GRBs in their models.
On the other hand, in our model, magnetic reconnection plays an important
role not only in the outflow but also in the accretion disk corona, and 
log-normal distribution is created
by the ``filtering effect'' in the reconnection associated with mass ejections 
in the accretion disk corona.

It is also known that soft gamma-ray repeaters (SGRs) are highly time-variable.  
It has been found that the distributions of the time intervals (waiting times) and total durations of bursts from SGRs show log-normal distributions (Hurley et al. 1994; G\"{o}\u{g}\"{u}\c{s} et al. 1999, 2000, 2001). Moreover, the intensity distribution of burst from SGRs shows the truncated log-normal distribution (Hurley et al. 1994), which is not clearly fitted to a log-normal distribution.
Therefore, the distributions of burst from SGRs are similar to those of GRBs.
We can possibly apply our flare/CME model to SGRs.

%% This section also includes several of the displayed math environments
%% mentioned in the Author Guide.

\acknowledgments

This CME catalog is generated and maintained by NASA and The Catholic University of America in cooperation with the Naval Research Laboratory. SOHO is a project of international cooperation between ESA and NASA.
This work was supported in part by the JSPS Japan-US Cooperation Science 
Program (principal investigators: K. Shibata and K. I. Nishikawa), by the Scientific Research Fund of the 
Ministry of Education, Science, and Culture (14540226), and by the JSPS Japan-UK Cooperation Science Program (principal investigators: K. Shibata and N. O. Weiss).

\epsscale{0.25}

\begin{figure}
\plotone{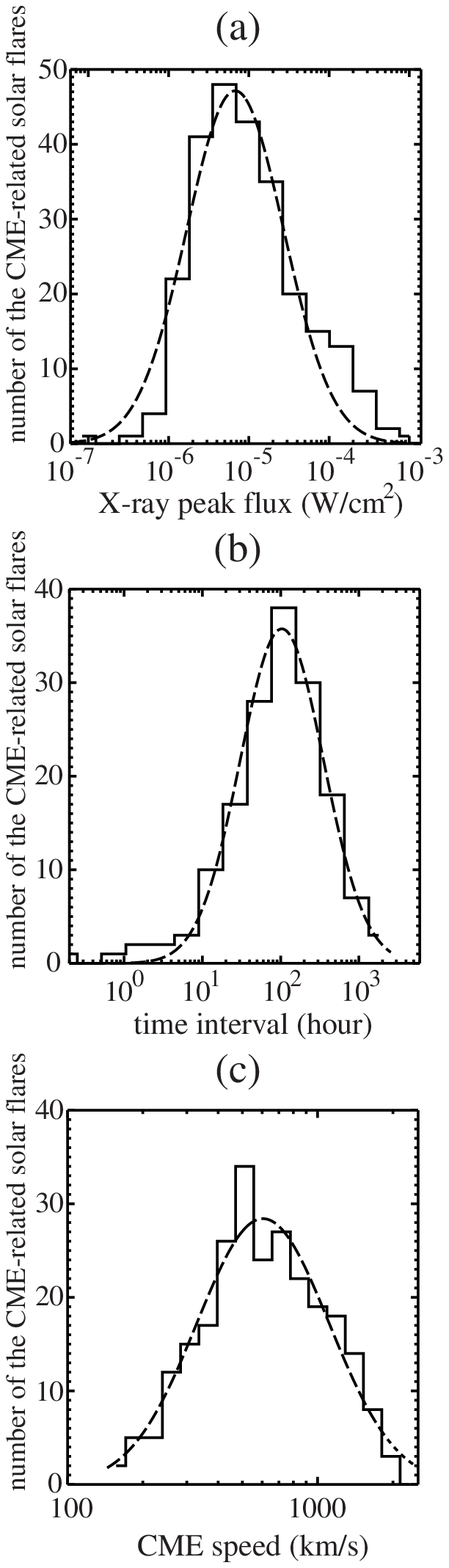}
\caption{Statistical distributions of the physical quantities of the CME-related solar flares. In each panel, the vertical axis shows the number of the CME-related solar flares in linear scale. The solid curves and the dashed curves show the histogram of the data and the best-fitting log-normal distribution, respectively. (a) The X-ray peak flux distribution of the CME-related solar flares. The horizontal axis shows the X-ray peak fluxes of the CME-related solar flares in unit of $\mathrm{W/cm^2}$ in logarithmic scale. The X-ray peak flux at the peak position of the number of the CME-related solar flares is $\sim 6.81 \times 10^{-6} \mathrm{W/cm^2}$. (b) The time interval distribution of the CME-related solar flares. The horizontal axis shows the time intervals of the CME-related solar flares in unit of an hour in logarithmic scale. The time interval at the peak position of the number of the CME-related solar flares is $\sim 118$ hours. (c) The CME speed distribution of the CME-related solar flares. The horizontal axis shows the speed of CMEs in unit of km/s in logarithmic scale. The speed at the peak position of the number of the CME-related solar flares is $\sim 605 \mathrm{km/s}$. 
\label{fig1}}
\end{figure}

\epsscale{0.7}

\begin{figure}
\plotone{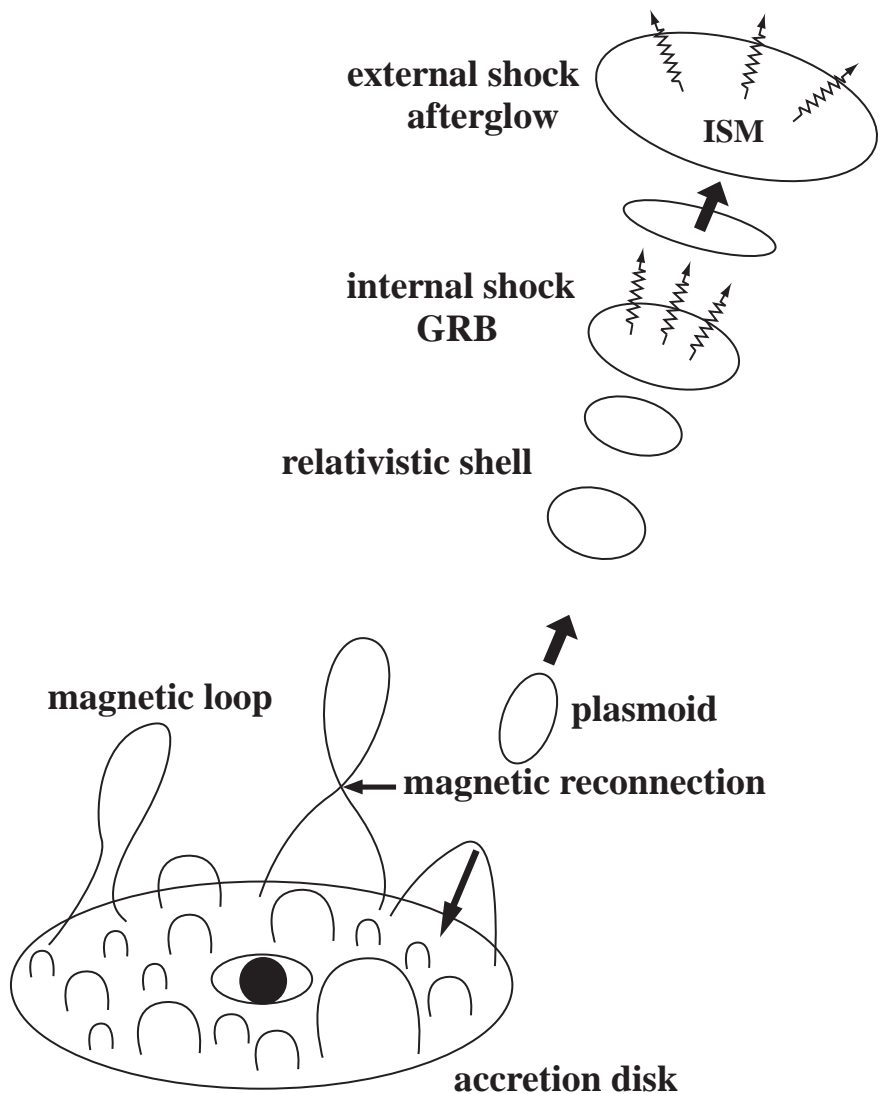}
\caption{The flare/CME model of the central engine of a gamma-ray burst (GRB). On the basis of the collapsar model and NS-NS(BH) merger model, we assume a system of a stellar mass black hole and a highly magnetized accretion disk. Hence, a lot of flares occur in magnetic loop on the disk via magnetic reconnection, and many plasmoids are produced and accelerated. If the flare is energetic, the plasmoid associated with the flare can be escaped from the magnetosphere of the disk. Various flares with different energy scales occur repeatedly on the disk, so that many plasmoids with various speeds are ejected from the central engine of a GRB. If the preceding plasmoid is slower than the following one, the collision between them occurs and an internal shock can be formed. The shell merges via the collision, and the merged shell still propagates and collides with interstellar matters (ISMs). Then external shocks are formed, which can be observed as afterglows.
\label{fig2}}
\end{figure}

%% If you are not including electonic art with your submission, you may
%% mark up your captions using the \figcaption command. See the 
%% No more than seven \figcaption commands are allowed per page, 
%% so if you have more than seven captions, insert a \clearpage 

\end{document}